\documentclass[fleqn,usenatbib]{mnras}

\usepackage{newtxtext,newtxmath}
\usepackage[T1]{fontenc}

\DeclareRobustCommand{\VAN}[3]{#2}
\let\VANthebibliography\thebibliography
\def\thebibliography{\DeclareRobustCommand{\VAN}[3]{##3}\VANthebibliography}

\usepackage{graphicx}
\usepackage{amsmath}
\usepackage{multirow}
\usepackage{threeparttable}

\title[Scintillation Arcs in PSR~B0355+54]{PSR~B0355+54: An Unexpected Compendium of Nearly All Scintillation Arc Morphologies}

\author[Liao et al.]{
Zhen-Long Liao,$^{1,2}$
Zhen Yan,$^{1,2,3}$\thanks{E-mail: yanzhen@shao.ac.cn}
Zhi-Qiang Shen,$^{1,2,3,4}$\thanks{E-mail: zshen@shao.ac.cn}
Rui Wang,$^{5}$
Xun Shi,$^{6}$
Ke-Jia Lee,$^{7,8}$\newauthor
Ju-Mei Yao,$^{2,9,10}$
Xiao-Wei Wang,$^{1,3}$
Zhi-Peng Huang,$^{11,12}$
Kuo Liu,$^{1,2,3}$
Yong-Hua Xu,$^{13}$\newauthor
Chu-Yuan Zhang,$^{1,2}$
Fan Yang $^{1,4}$
and Yang-Yang Lin$^{1}$
\\
$^{1}$Shanghai Astronomical Observatory, Chinese Academy of Sciences, Shanghai 200030, China\\
$^{2}$School of Astronomy and Space Sciences, University of Chinese Academy of Sciences, Beijing 100049, China\\
$^{3}$State Key Laboratory of Radio Astronomy and Technology, Shanghai Astronomical Observatory, Chinese Academy of Sciences, Shanghai 200030, China\\
$^{4}$School of Physical Science and Technology, ShanghaiTech University, Shanghai 201210, China\\
$^{5}$School of Information Engineering, Sanming University, Sanming 365004, China\\
$^{6}$South-Western Institute for Astronomy Research, Yunnan University, Kunming 650500, China\\
$^{7}$Department of Astronomy, School of Physics, Peking University, Beijing 100871, China\\
$^{8}$Kavli Institute for Astronomy and Astrophysics, Peking University, Beijing 100871, China\\
$^{9}$Xinjiang Astronomical Observatory, Chinese Academy of Sciences, Xinjiang 830011,China\\
$^{10}$State Key Laboratory of Radio Astronomy and Technology, Xinjiang Astronomical Observatory, Chinese Academy of Sciences, Xinjiang 830011, China\\
$^{11}$School of Optoelectronic Information Engineering, Hubei University of Education, Wuhan 430205, China\\
$^{12}$Institute of Astronomy and High Energy Physics, Hubei University of Education, Wuhan 430205, China\\
$^{13}$Yunnan Observatories, Chinese Academy of Sciences, Kunming 650011, China
}

\date{Accepted XXX. Received YYY; in original form ZZZ}
\pubyear{2026}

\begin{document}
\label{firstpage}
\pagerange{\pageref{firstpage}--\pageref{lastpage}}
\maketitle

\begin{abstract}
Scintillation arcs are sensitive probes of multipath propagation, offering a window into the distribution of scattering material in the interstellar medium. Contrary to the common observational finding that individual pulsars exhibit only a limited range of scintillation arc morphologies, we report that PSR~B0355+54, observed simultaneously at 2.25/8.60~GHz with the Shanghai Tianma Radio Telescope, displays nearly the full range of known arc patterns, many of which evolve rapidly. Across 54 observations from 2020 to 2026, we detect scintillation arcs in 53/53 usable observations at 2.25~GHz and 33/49 at 8.60~GHz. At 2.25~GHz, the secondary spectra exhibit five broad morphological types: forward arcs, inverted arclets, diffuse arcs, diffuse power, and irregular structures, with prominent displaced arclets appearing in several observations. In contrast, the 8.60-GHz spectra are generally dominated by faint forward arcs. Despite these marked morphological differences, the scaled curvatures and their annual variations agree between the two frequencies, showing that both trace the same dominant scattering screen at a fractional distance of $s=0.069^{+0.015}_{-0.014}$ near the pulsar, further supporting a possible association with the \ion{H}{ii} region Sh~2-205. Simulations show that this diversity can arise from variations in the scattered-image distribution within this screen, while brightness reconstructions at 2.25~GHz resolve localized structure down to $\sim0.02~\mathrm{mas}$. Tracking four arclets over 10~h at 8.60~GHz, including two displaced arclets, reveals a common evolutionary trend consistent with the line of sight sweeping across small-scale scattering structure, helping explain the rapid arc evolution.
\end{abstract}



\section{Introduction}

The ionized component of the interstellar medium (ISM) is widely distributed throughout the Galaxy and provides important information on thermal and dynamical processes in the ISM. However, much of this material is diffuse and low-density, making its spatial distribution and electron density structure difficult to characterize in detail. Owing to their compact emission regions, pulsars can exhibit prominent interstellar scintillation (ISS) when their radio waves undergo multipath scattering and interference caused by electron density inhomogeneities in the ionized ISM. This makes pulsar ISS an important probe of the distribution of scattering material, electron density fluctuations, and propagation properties of the ionized ISM \citep{Rickett1990,Narayan1992}. Pulsar ISS has been studied since the early years of pulsar astronomy \citep{Lyne1968,Scheuer1968}. While refractive interstellar scintillation (RISS) accounts for broadband flux density variations on time-scales of days to months, diffractive interstellar scintillation (DISS) drives intensity modulation in the dynamic spectrum (i.e. intensity as a function of time and frequency) typically over narrow frequency scales ($\sim$kHz--MHz) and short time-scales ($\sim$minutes--hours) \citep{Rickett1969,Rickett1984,Rickett1990}. Traditional ISS studies quantify DISS using the decorrelation bandwidth $\Delta\nu_{\mathrm{d}}$ and scintillation time-scales $\Delta\tau_{\mathrm{d}}$, usually measured from the two-dimensional (2D) autocorrelation function (ACF) of the dynamic spectrum. These parameters have been widely applied to study the distribution of scattering material along the line of sight (LOS) and to estimate pulsar transverse velocities from scintillation speeds \citep{Cordes1986,Gupta1995,Cordes1998}.

A further advance came from the discovery of parabolic scintillation arcs in the secondary spectrum (i.e. the 2D Fourier power spectrum of the dynamic spectrum), revealing organized interference structures within the apparently stochastic intensity variations of pulsar ISS \citep{Stinebring2001}. These arcs can constrain the locations of dominant scattering screens and provide sensitive geometric probes of multipath propagation, enabling studies of localized plasma structures on au and sub-au scales \citep{Walker2004,Cordes2006}. Numerous observations have since shown that scintillation arcs are common in high-quality secondary spectra and exhibit rich morphological diversity across the pulsar population, whereas observations of individual pulsars generally reveal a more restricted set of features \citep[e.g.][]{Hill2005,Wang2018,Stinebring2022,Main2023a,Main2023b,Ocker2024}. Such morphological differences can be sensitive to the geometry and angular brightness distribution of the scattered image, and may provide clues to the structure and anisotropy of the intervening plasma \citep{Brisken2010,Stinebring2019,Rickett2021}.

Several approaches have been developed to constrain this geometry. With spatial baseline information, very long baseline interferometry (VLBI) and two-station intensity scintillometry have directly resolved scattered images at angular resolutions of $\sim0.1~\mathrm{mas}$, revealing detailed angular structures that can be highly anisotropic \citep{Brisken2010,Sprenger2026}. In suitable cases, techniques such as interstellar holography and the $\theta$--$\theta$ transform have been used to recover the scattered wavefield or projected brightness distribution from scintillation data \citep{Walker2008,Sprenger2021,Baker2022,Zhu2023,Chen2025}. Qualitative simulations can also help interpret the observed arc morphologies by exploring how different scattered image distributions affect the secondary spectrum \citep{Mall2022}. These geometric constraints can then be compared with independently traced ISM structures to investigate the physical origin of the scattering. Several studies have associated scintillation screens with the Local Bubble, \ion{H}{ii} regions, supernova remnants, and pulsar bow shocks \citep[e.g.][]{Yao2021,Mall2022,Wang2025,Reardon2025}. However, such associations have been reported so far for only a limited number of scattering screens.

PSR~B0355+54 is located at $1.0^{+0.2}_{-0.1}~\mathrm{kpc}$ \citep{Verbiest2012}, with proper motion $[\mu_\alpha,\mu_\delta]=[9.20\pm0.18,\ 8.17\pm0.39]~\mathrm{mas~yr^{-1}}$ \citep{Chatterjee2004}. Its scintillation arcs were first reported at 1.4~GHz with the Green Bank Telescope (GBT) \citep{Stinebring2007}. Subsequent 2.25-GHz monitoring with the Kunming 40-m telescope (KM40) revealed variations in scintillation bandwidth, time-scale, and arc curvature, suggesting changes in the scattering screens or ray paths \citep{Xu2018}. Later 2.25-GHz observations with the Jiamusi 66-m telescope (JM66) detected inverted arclets evolving along the main arc, yielding an angular velocity consistent with the VLBI proper motion \citep{Wang2018}. More recent 1.4-GHz observations with FAST revealed four arcs with distinct curvatures, indicating multiple scattering structures along the LOS \citep{Ocker2024}. However, several candidate scattering screens have been proposed toward PSR~B0355+54, with no single scenario firmly established. A nearby screen tentatively associated with the Local Bubble boundary has been suggested \citep{Xu2018}, while an alternative interpretation places a screen close to the pulsar \citep{Wang2018}. The FAST curvature measurements retain a twofold degeneracy between near-observer and near-pulsar screen solutions, with the near-pulsar solution for the lowest-curvature arc allowing a possible association with the putative pulsar bow shock \citep{Ocker2024}. Meanwhile, the LOS intersects the edge of the H~{\sc ii} region Sh~2-205, with distance estimates around 1~kpc \citep{Romero2008,Foster2015,Dong2026}. This region has been considered a plausible scattering contributor toward PSR~B0355+54 \citep{Mitra2003,Ocker2024,Ocker2024HII}. These candidate environments may host distinct scattering structures, but their specific associations with the observed arcs remain uncertain.

As early as 2018, pioneering simultaneous 2.25/8.60~GHz pulsar observations were conducted with the Shanghai Tianma Radio Telescope (TMRT), a fully steerable 65-m antenna with high sensitivity, particularly at higher frequencies \citep{Yan2018}. This simultaneous dual-frequency observing capability has subsequently been applied more broadly, yielding a series of results in areas such as constraining pulsar radiation properties and understanding ISM properties \citep{Wang2024,Wang2025,Wang2026,Huang2026}. In this paper, we report long-term simultaneous 2.25/8.60~GHz observations of PSR~B0355+54 using the TMRT, which enable us to trace the evolution of its scintillation arcs and constrain the geometry of the dominant scattering screen. Section~\ref{sec:2} describes the observations, Section~\ref{sec:3} presents the data analysis and observational results, Section~\ref{sec:4} contains the discussion, and Section~\ref{sec:5} summarizes the conclusions.

\section{Observations}
\label{sec:2}

PSR~B0355+54 was observed simultaneously at 2.25 and 8.60~GHz using the TMRT, with the same setup as in previous observations \citep{Yan2018}. The effective frequency coverages were 2.2--2.3 and 8.2--9.0~GHz, respectively. The dual-frequency signals were sampled and pre-processed by the Digital Backend System (DIBAS), which provided frequency resolutions (sub-channel widths) of 0.97~MHz at 2.25~GHz and 1.95~MHz at 8.60~GHz, respectively. Using its online incoherent dedispersion mode, the signals were dedispersed to correct for interstellar dispersion and then folded into 30-s subintegrations. The resulting data were recorded in standard PSRFITS format for subsequent processing with commonly used pulsar analysis software \citep{Psrchive2004,Yan2015,Yan2024}. Of the 54 observations, 53 provided usable data at 2.25~GHz and 49 at 8.60~GHz.

\begin{table*}
    \centering
    \caption{Observational parameters and scintillation measurements for the first five observations: Modified Julian Date (MJD), observation length ($T$), observing frequency ($\nu$), scintillation time-scale ($\Delta\tau_{\mathrm{d}}$), decorrelation bandwidth ($\Delta\nu_{\mathrm{d}}$), drift rate ($\mathrm{d}t/\mathrm{d}\nu$), DISS modulation index ($m$), scaled pseudo-curvature ($\eta_{\mathrm{iss,scaled}}$), scaled arc curvature ($\eta_{\mathrm{scaled}}$), curvature ratio ($R=\eta_{\mathrm{iss,scaled}}/\eta_{\mathrm{scaled}}$), and arc morphology. The morphology classifications are described in Appendix~\ref{app:morphology}, and the complete measurements for all observations are provided in Table~S1 of the Supporting Information.}
    \label{tab:exparameters}
    \setlength{\tabcolsep}{1.5pt}
    \medmuskip=2mu plus 1mu minus 1mu\relax
    \begin{tabular*}{\textwidth}{@{\extracolsep{\fill}}ccccccccccc@{}}
        \hline
        MJD
        & $T$
        & $\nu$
        & $\Delta\tau_{\mathrm{d}}$
        & $\Delta\nu_{\mathrm{d}}$
        & $\mathrm{d}t/\mathrm{d}\nu$
        & $m$
        & $\eta_{\mathrm{iss,scaled}}$
        & $\eta_{\mathrm{scaled}}$
        & $R$
        & Arc morphology \\
        --
        & (s)
        & (GHz)
        & (min)
        & (MHz)
        & (s MHz$^{-1}$)
        & --
        & $(\mathrm{m}^{-1}\,\mathrm{mHz}^{-2})$
        & ($\mathrm{m}^{-1}\,\mathrm{mHz}^{-2}$)
        & --
        & -- \\
        \hline

        \multirow{2}{*}{59131.89}
        & \multirow{2}{*}{4450}
        & 2.25 & $2.11 \pm 0.10$ & $2.36 \pm 0.12$
        & $14.17 \pm 1.21$ & 0.84
        & $459.20 \pm 49.39$
        & $271.54 \pm 14.41$
        & $1.69 \pm 0.20$
        & M3 \\
        & & 8.60 & $6.85 \pm 1.28$ & $221.40 \pm 41.50$
        & $0.32 \pm 0.09$ & 0.75
        & -- & -- & -- & -- \\

        \multirow{2}{*}{59132.82}
        & \multirow{2}{*}{7997}
        & 2.25 & $2.73 \pm 0.13$ & $3.30 \pm 0.16$
        & $12.88 \pm 0.96$ & 0.93
        & $549.75 \pm 58.75$
        & $266.74 \pm 9.88$
        & $2.06 \pm 0.23$
        & M3 \\
        & & 8.60 & $11.32 \pm 2.17$ & $290.66 \pm 55.63$
        & $0.50 \pm 0.13$ & 0.53
        & --
        & $258.51 \pm 8.93$
        & --
        & M1 \\

        \multirow{2}{*}{59135.79}
        & \multirow{2}{*}{7667}
        & 2.25 & $2.62 \pm 0.15$ & $3.86 \pm 0.20$
        & $0.00 \pm 0.25$ & 0.78
        & $432.88 \pm 54.40$
        & --
        & --
        & M4 \\
        & & 8.60 & $10.72 \pm 2.32$ & $786.83 \pm 170.18$
        & $-0.54 \pm 0.17$ & 0.75
        & -- & -- & -- & -- \\

        \multirow{2}{*}{59136.81}
        & \multirow{2}{*}{11996}
        & 2.25 & $5.55 \pm 0.52$ & $6.06 \pm 0.43$
        & $-7.46 \pm 0.86$ & 0.91
        & $1237.27 \pm 247.91$
        & $263.17 \pm 16.00$
        & $4.70 \pm 0.98$
        & M5 \\
        & & 8.60 & $23.65 \pm 5.42$ & $414.05 \pm 94.76$
        & $-0.44 \pm 0.14$ & 0.66
        & --
        & $281.41 \pm 12.04$
        & --
        & M1 \\

        \multirow{2}{*}{59138.80}
        & \multirow{2}{*}{5893}
        & 2.25 & $2.30 \pm 0.12$ & $2.74 \pm 0.13$
        & $25.75 \pm 1.75$ & 0.89
        & $469.96 \pm 53.87$
        & $258.51 \pm 16.37$
        & $1.82 \pm 0.24$
        & M3 \\
        & & 8.60 & $19.84 \pm 4.84$ & $219.64 \pm 53.57$
        & $0.58 \pm 0.20$ & 0.58
        & -- & -- & -- & -- \\

        $\cdots$ & $\cdots$ & $\cdots$ & $\cdots$ & $\cdots$
        & $\cdots$ & $\cdots$
        & $\cdots$
        & $\cdots$
        & $\cdots$
        & $\cdots$ \\
        \hline
    \end{tabular*}
\end{table*}

\section{Data analysis and results}
\label{sec:3}
\subsection{Dynamic spectra and 2D ACF}
\label{sec:acf}
The dynamic spectra were constructed after manually flagging strong radio frequency interference (RFI) using the \texttt{PSRCHIVE} utility \texttt{pazi}\footnote{\url{http://psrchive.sourceforge.net/}} \citep{Psrchive2004,Psrchive2012}. For each unflagged subintegration and frequency channel, the pulse intensity was calculated from the difference between the mean on-pulse and off-pulse intensities and normalized by the time-averaged off-pulse intensity to correct for the bandpass response. Residual RFI in the dynamic spectra was manually removed, and missing pixels were filled by linear interpolation. The fractional intensity variations were characterized by the modulation index $m=\sigma_I/\langle I\rangle$, where $\sigma_I$ and $\langle I\rangle$ are the standard deviation and mean intensity of the dynamic spectrum, respectively.

We measured the scintillation parameters from the ACFs using the package \texttt{scintools}\footnote{\url{https://github.com/danielreardon/scintools}} \citep{Reardon2020,Reardon2023}, following our previous work \citep{Wang2025,Wang2026}. Because broad scintillation patches are only partially sampled in frequency in some observations, we modified the fitting procedure by first fitting the zero-frequency-lag cut of the ACF to obtain a more reliable estimate of $\Delta\tau_{\mathrm d}$, and then fitting the full 2D ACF to determine $\Delta\nu_{\mathrm d}$ and the drift rate $\mathrm{d}t/\mathrm{d}\nu$. The quoted uncertainties include both fitting and finite-scintle errors \citep{Cordes1986b}. The fitted ACF parameters for the first five observations are listed in columns 4--6 of Table~\ref{tab:exparameters} (the complete measurements and corresponding dynamic spectra and 2D ACFs are provided in Table~S1 and Fig.~S1 of the Supporting Information, respectively). The available observing bandwidth is approximately 100~MHz at 2.25~GHz and 800~MHz at 8.60~GHz. When broad scintillation patches are not fully sampled within these bands, the fitted $\Delta\nu_{\mathrm d}$ can approach or even exceed the available bandwidth. This occurs in most of the 8.60-GHz observations, for which the fitted values should therefore be regarded as indicative estimates rather than well-constrained measurements.

\subsection{Secondary spectra and curvature fitting}

\begin{figure*}
    \centering
    \includegraphics[]{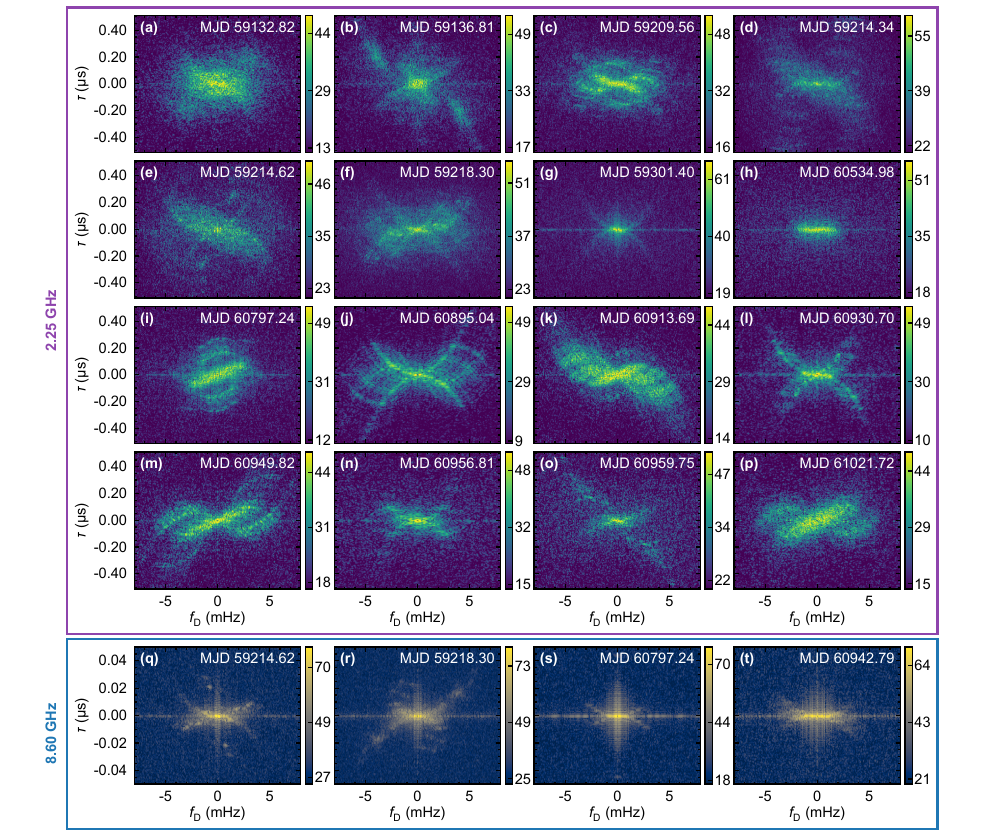}
    \caption{Representative secondary spectra of PSR~B0355+54 at 2.25~GHz [panels (a)--(p), purple frame] and 8.60~GHz [panels (q)--(t), blue frame]. Colour bars denote the logarithmic power defined in equation~\eqref{eq:SS}, with independent display ranges applied to each panel to better reveal the detailed spectral features.}
    \label{fig:sspec}

    \noindent\begin{minipage}{\textwidth}
    \textbf{Alt text:} Twenty representative secondary spectra arranged in two frequency groups. The sixteen 2.25-GHz panels show diverse morphologies, including forward arcs, inverted arclets, diffuse arcs, diffuse power, and irregular structures, with displaced arclets visible in several panels. The four 8.60-GHz panels are generally dominated by faint forward arcs.
    \end{minipage}
\end{figure*}

To probe the scattering geometry, we computed the secondary spectrum from the 2D Fourier transform of the mean-subtracted dynamic spectrum, with a Hamming window function applied to the outer 10 per cent along both the time and frequency dimensions to suppress spectral leakage. The secondary spectrum is defined as
\begin{equation}
    S(f_{\mathrm{D}},\tau)=10\log_{10}\left|\widetilde{I}(f_{\mathrm{D}},\tau)\right|^2,
    \label{eq:SS}
\end{equation}
where $\widetilde{I}(f_{\mathrm{D}},\tau)$ is the 2D Fourier transform of the processed dynamic spectrum, and $f_{\mathrm{D}}$ and $\tau$ represent the Doppler frequency and differential delay, respectively.

Scintillation arcs are detected in all 53 usable 2.25-GHz observations and in 33 of the 49 usable 8.60-GHz observations. Representative secondary spectra are presented in Fig.~\ref{fig:sspec}, with the full set provided in Fig.~S1 of the Supporting Information. Using the classifications defined in Appendix~\ref{app:morphology}, the 2.25-GHz secondary spectra show substantial morphological diversity, with examples of (M1) forward arcs in panels (g) and (l), (M2) inverted arclets in panels (c) and (m), (M3) diffuse arcs in panels (a) and (d), (M4) diffuse power in panel (h), and (M5) irregular structures in panels (b) and (o). Prominent displaced arclets also appear as distinct substructures in several observations (e.g. panels (c) and (i)). Panels (c)--(f) also show rapid morphological changes on time-scales of days or even hours. Although individual morphological features and temporal variations have been reported in other pulsars, such rich morphological diversity and rapid transitions within a single source remain exceptional \citep[e.g.][]{Brisken2010,Mall2022,Sprenger2022,Chen2025}. In contrast, the 8.60-GHz spectra are generally dominated by faint forward arcs (e.g. panels (q) and (s)), with only weak displaced arclets seen in a few observations (e.g. panel (r)). Direct comparison of the simultaneous observations shows that the two bands sometimes exhibit similar overall orientations and asymmetric extensions, as seen in panels (f) and (r), but can differ markedly in morphology in other observations, as in panels (i) and (s). The morphology classifications for the first five observations are listed in Table~\ref{tab:exparameters}, with the complete results provided in Table~S1 of the Supporting Information. Possible explanations for these morphologies are explored further through thin-screen simulations in Section~\ref{sec:modelling}.

There are several approaches to measuring arc curvature, including searches for power concentrated along parabolic features in the secondary spectrum \citep{Bhat2016,Xu2018,Reardon2020} and coordinate transformations such as the $\theta$--$\theta$ transform, which maps parabolic interference structures into linear features \citep{Sprenger2021,Baker2022}. Given that many of our observations show only faint forward arcs, particularly at 8.60~GHz, while some exhibit complex structures with diffuse or displaced features, identifying clear linear structures in the transformed space can be challenging. Nevertheless, individual parabolic features can still be identified in most secondary spectra, even when they are not clearly defined throughout their extent or are surrounded by other structures (e.g. panel (e) of Fig.~\ref{fig:sspec}). We therefore use a 2D Hough transform to measure arc curvature consistently across the data set by identifying these features based on their power contrast with the local background.

Specifically, we search for parabolic features described by
\begin{equation}
    \tau=\eta f_{\mathrm{D}}^{2}-2\eta b f_{\mathrm{D}},
\end{equation}
where $\eta$ is the arc curvature and $b$ specifies the Doppler offset of the parabola apex. For each pair of $(\eta,b)$, a normalized statistic similar in form to that of \citet{Xu2018} is calculated by comparing the power along the corresponding parabola with its local background. The background region is defined by expanding the parabolic sampling mask and excluding the original mask. We then identify the dominant peak of this statistic in the $(\eta,b)$ plane and fit it with a rotated 2D Gaussian. The fitted centroid gives the best-fitting arc curvature $\eta$, while the projected $1\sigma$ width of the Gaussian along the $\eta$ axis is adopted as an empirical estimate of the curvature uncertainty. An example of the Hough measurement and peak fitting is shown in Fig.~S2 of the Supporting Information.

To compare the curvature measurements at the two observing frequencies, we calculate the scaled curvature as $\eta_{\mathrm{scaled}}=\eta\nu^2/c\simeq3.3\times10^3\eta\nu^2$, where $\nu$ is the observing frequency and $c$ is the speed of light \citep{Ocker2024}. Here, $\eta$, $\nu$, and $\eta_{\mathrm{scaled}}$ are expressed in $\mathrm{s^3}$, GHz, and $\mathrm{m^{-1}\,mHz^{-2}}$, respectively. The resulting values are generally consistent between the two frequencies within their uncertainties, as expected for the $\eta\propto\nu^{-2}$ scaling, suggesting that the two bands may probe a common scattering geometry.

\section{Discussions}
\label{sec:4}

\subsection{The location of the scattering screen}
\label{sec:location}

For an isolated pulsar under the thin-screen approximation, the measured arc curvature can exhibit an annual modulation owing to the Earth's orbital motion \citep{Cordes1998,Xu2023}. For a screen at distance $D_{\mathrm{scr}}$ from the observer and a pulsar at distance $D_{\mathrm{p}}$, we use the fractional screen distance as $s=1-D_{\mathrm{scr}}/D_{\mathrm{p}}$. The corresponding effective distance and transverse velocity are
\begin{align}
    D_{\mathrm{eff}}
    &=D_{\mathrm{p}}\frac{1-s}{s},\\
    \boldsymbol{V}_{\mathrm{eff}}
    &=\boldsymbol{V}_{\oplus}
    +\frac{1-s}{s}\boldsymbol{V}_{\mathrm{p}}
    -\frac{1}{s}\boldsymbol{V}_{\mathrm{scr}},
\end{align}
where $\boldsymbol{V}_{\mathrm{p}}$, $\boldsymbol{V}_{\oplus}$, and $\boldsymbol{V}_{\mathrm{scr}}$ are the transverse velocities of the pulsar, Earth, and scattering screen, respectively. For anisotropic scattering, the scaled curvature is
\begin{equation}
    \eta_{\mathrm{scaled}}
    =\frac{D_{\mathrm{eff}}}{2V_{\mathrm{eff}}^{2}\cos^{2}\Delta}
    =\frac{D_{\mathrm{eff}}}
    {2\left[
    V_{\mathrm{eff},\alpha}\sin\psi_0
    +
    V_{\mathrm{eff},\delta}\cos\psi_0
    \right]^2},
    \label{eq:annual_curvature}
\end{equation}
where $V_{\mathrm{eff},\alpha}$ and $V_{\mathrm{eff},\delta}$ are the effective velocity components along increasing right ascension and declination, respectively. The scattering axis position angle $\psi_0$ is measured from the $+\delta$ direction toward the $+\alpha$ direction, and $\Delta$ is the angle between $\boldsymbol{V}_{\mathrm{eff}}$ and the scattering axis. In this model, the curvature depends on the component of the effective velocity along the scattering axis. For isotropic scattering, the full transverse velocity is used, formally corresponding to $\cos^2\Delta=1$ in equation~\eqref{eq:annual_curvature}.

\begin{figure}
    \centering
    \includegraphics[]{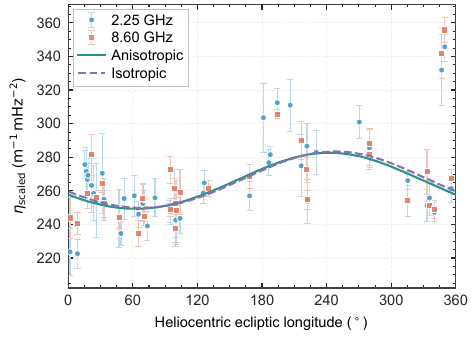}
    \caption{Annual variation of the scaled scintillation arc curvature, $\eta_{\mathrm{scaled}}$, with heliocentric ecliptic longitude. Blue circles and orange squares show the 2.25- and 8.60-GHz measurements, while the solid and dashed curves show the anisotropic and isotropic joint-fit models. The two outliers near $345^\circ$ correspond to MJDs 60927 and 60930, discussed in the text.}
    \label{fig:eta}
    \noindent\begin{minipage}{\columnwidth}
    \textbf{Alt text:} Scaled scintillation arc curvature as a function of heliocentric ecliptic longitude for observations at 2.25 and 8.60~GHz. Measurements at the two frequencies show a similar overall annual trend, and the anisotropic and isotropic joint-fit curves nearly overlap throughout the year.
    \end{minipage}
\end{figure}

\begin{table}
    \centering
    \caption{Model parameters derived from the annual curvature fits.}
    \label{tab:annual_fit}
    \setlength{\tabcolsep}{3pt}
    \renewcommand{\arraystretch}{1.25}
    \begin{tabular}{@{}clcccc@{}}
        \hline
        Data & Model
        & $s$
        & $V_{\mathrm{scr},\alpha}$
        & $V_{\mathrm{scr},\delta}$
        & $\psi_0$ \\
        (GHz) & -- & --
        & ($\mathrm{km\,s^{-1}}$)
        & ($\mathrm{km\,s^{-1}}$)
        & ($^\circ$) \\
        \hline

        \multirow{2}{*}{2.25}
        & Anisotropic
        & $0.075^{+0.017}_{-0.016}$
        & $-18.45^{+11.11}_{-9.96}$
        & $1.53^{+16.25}_{-15.70}$
        & $83.70^{+10.81}_{-15.49}$ \\
        & Isotropic
        & $0.075^{+0.019}_{-0.016}$
        & $-17.30^{+8.18}_{-8.07}$
        & $16.29^{+11.80}_{-13.73}$
        & -- \\

        \multirow{2}{*}{8.60}
        & Anisotropic
        & $0.052^{+0.018}_{-0.015}$
        & $7.31^{+14.09}_{-11.78}$
        & $0.15^{+16.01}_{-14.95}$
        & $81.75^{+13.66}_{-16.49}$ \\
        & Isotropic
        & $0.058^{+0.019}_{-0.016}$
        & $-7.08^{+9.78}_{-9.29}$
        & $13.90^{+13.18}_{-14.31}$
        & -- \\

        \multirow{2}{*}{Joint}
        & Anisotropic
        & $0.069^{+0.015}_{-0.014}$
        & $-16.17^{+10.37}_{-11.18}$
        & $-3.37^{+13.70}_{-18.89}$
        & $86.06^{+10.01}_{-10.44}$ \\
        & Isotropic
        & $0.067^{+0.015}_{-0.014}$
        & $-15.41^{+9.02}_{-6.82}$
        & $21.66^{+10.83}_{-12.41}$
        & -- \\

        \hline
    \end{tabular}
\end{table}

We fitted the scaled curvatures measured at 2.25 and 8.60~GHz separately and jointly using both isotropic and anisotropic models, accounting for the uncertainty in the pulsar distance ($D_{\mathrm{p}}=1.0^{+0.2}_{-0.1}~\mathrm{kpc}$). The posterior distributions were sampled using the \texttt{emcee} Markov chain Monte Carlo algorithm \citep{mcmc2013}, adopting a Student-$t$ likelihood with four degrees of freedom to reduce sensitivity to outliers. The resulting parameter constraints are listed in Table~\ref{tab:annual_fit}, with the joint fits shown in Fig.~\ref{fig:eta} and the corresponding posterior distributions provided in Fig.~S3 of the Supporting Information. Two observations at MJDs~60927 and 60930 show noticeable departures from the otherwise smooth annual variation. Similar curvature variations have been reported for PSR~B0355+54 and may reflect changes in the dominant scattering screen or ray path \citep{Xu2018}.

All fits yield consistent screen locations, with similar scattering axis orientations in the anisotropic models, while the screen velocity components are less well constrained. These results suggest that the scintillation observed at 2.25 and 8.60~GHz toward PSR~B0355+54 is dominated by the same scattering screen. Given the uncertainties in the individual frequency curvature measurements, we focus on the joint fits below.

The joint fits to both frequencies give nearly identical fractional screen distances: $s=0.069^{+0.015}_{-0.014}$ for the anisotropic model and $0.067^{+0.015}_{-0.014}$ for the isotropic model. Both correspond to $D_{\mathrm{scr}}=0.93^{+0.19}_{-0.09}~\mathrm{kpc}$, with the uncertainty dominated by that of the pulsar distance. In the anisotropic model, $\psi_0\simeq86^\circ$ makes the velocity projected along the scattering axis dominated by its $\alpha$ component, leaving $V_{{\mathrm{scr}},\delta}$ only weakly constrained. The effective velocity direction in the isotropic fit varies from about $74^\circ$ to $76^\circ$ over the year, close to the anisotropic axis orientation within uncertainties. Consequently, the two models predict very similar curvature variations. For reference, the anisotropic and isotropic joint fits yield reduced $\chi^2$ values of 7.45 and 7.35, respectively, showing nearly identical levels of residual scatter and no clear preference between the two geometries. The prominent inverted arclets in the secondary spectra may be more consistent with an anisotropic interpretation of the dominant scattering screen. We therefore focus on the anisotropic solution, with $s=0.069^{+0.015}_{-0.014}$ and $\psi_0=86.06^{+10.01}_{-10.44}\,^\circ$, in the following analysis.

To further test this scattering interpretation, we compare the measured arc curvature with the ACF-derived pseudo-arc curvature, as the two quantities are expected to agree for a single thin screen with anisotropic scattering \citep{Stinebring2022}. We calculate the pseudo-curvature as
\begin{equation}
\eta_{\mathrm{iss}}=
\frac{2\pi(\ln 2)^{6/5}\Delta\tau_{\mathrm d}^{2}}
{\Delta\nu_{\mathrm d}},
\end{equation}
where the factor $(\ln 2)^{6/5}$ converts the $1/e$ ACF time-scale used here to the half-maximum time-scale adopted by \citet{Stinebring2022}, assuming a Kolmogorov exponent of $5/3$. Given the limited observing bandwidth discussed in Section~\ref{sec:acf}, we calculate $\eta_{\mathrm{iss}}$ only for the 2.25-GHz observations with fitted $\Delta\nu_{\mathrm d}$ values smaller than half of the observing bandwidth, for which $\Delta\nu_{\mathrm d}$ is more reliably constrained. We compare the two curvature estimates using $R=\eta_{\mathrm{iss,scaled}}/\eta_{\mathrm{scaled}}$, where $\eta_{\mathrm{iss,scaled}}$ is scaled in the same way as $\eta_{\mathrm{scaled}}$. Although the $R$ values listed in Table~\ref{tab:exparameters} lie above unity, the complete set of valid measurements shows no clear systematic offset from $R=1$, with a median of 1.01 and a 16th--84th percentile range of 0.74--1.50 (see Table~S1 of the Supporting Information for the complete measurements and Fig.~S4 for a visual comparison). Considerable scatter remains, and some of the larger deviations may arise from biases in the ACF parameters in the presence of periodic or criss-cross scintillation patterns \citep{Yao2026}. This overall agreement therefore provides an additional consistency check on the dominant anisotropic-screen interpretation adopted in the annual curvature analysis.

\begin{figure*}
    \centering
    \includegraphics[]{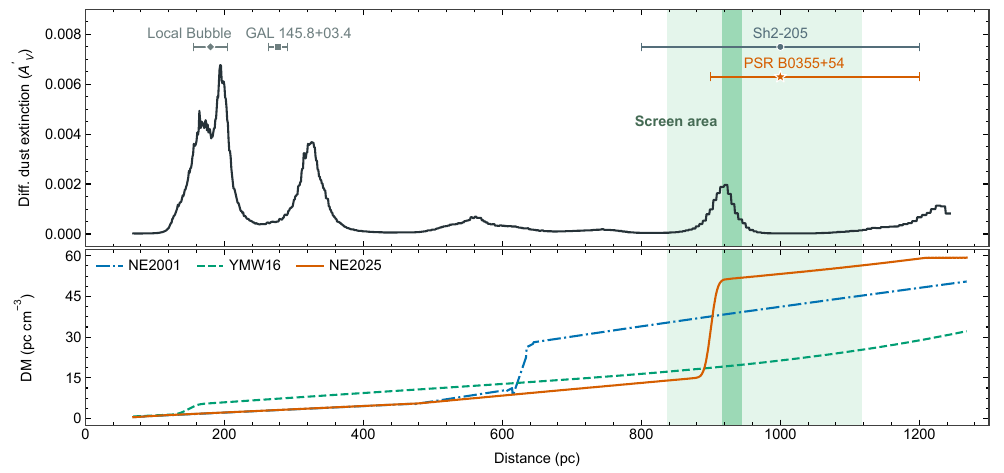}
    \caption{Comparison of the inferred scattering screen distance with the differential dust extinction (upper panel) and Galactic DM--distance relations (lower panel) along the LOS. Horizontal error bars in the upper panel indicate the distance estimates and uncertainties for known ISM structures along the LOS and PSR~B0355+54. The light and dark shaded regions denote the inferred screen distance ranges when the pulsar distance uncertainty is included and when $D_{\mathrm{p}}=1.0~\mathrm{kpc}$ is fixed, respectively.}
    \label{fig:asco}
    \noindent\begin{minipage}{\textwidth}
    \textbf{Alt text:} Two-panel comparison of the inferred scattering-screen distance with dust extinction and Galactic DM--distance relations along the LOS. The screen region overlaps a dust enhancement and the distance range of Sh~2-205, while the NE2025 relation shows a sharp DM increase at a similar distance.
    \end{minipage}
\end{figure*}

To investigate the possible physical origin of this dominant screen, we compare its inferred distance with the differential dust extinction from the 3D dust map \citep{Edenhofer2023}, the DM--distance relations predicted by the NE2001, YMW16, and NE2025 Galactic electron density models using \texttt{PyGEDM}\footnote{\url{https://apps.datacentral.org.au/pygedm/}} \citep{ne2001,ymw2016,ne2025,pygedm}, and known ISM structures along the LOS \citep{Romero2008,Yan2021,Neill2024}, as shown in Fig.~\ref{fig:asco}. The inferred screen lies near a prominent dust enhancement and is consistent with distance estimates for Sh~2-205 \citep{Romero2008,Foster2015,Dong2026}. Among the electron density models, NE2025 shows a pronounced rise in DM over a similar distance range, reflecting enhanced electron density along the LOS. The screen location is also compatible with the near-pulsar solution ($0.031\lesssim s\lesssim0.08$) for the FAST arc with a comparable scaled curvature \citep{Ocker2024}. These comparisons support a possible association of the dominant scattering screen with Sh~2-205.

\subsection{Scattering simulations of observed arc morphologies}
\label{sec:modelling}

\begin{figure}
    \centering
    \includegraphics[]{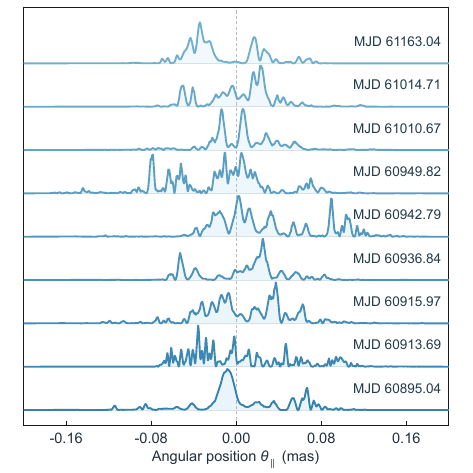}
    \caption{Reconstructed 1D brightness distributions of the scattered image for selected 2.25-GHz observations along the dominant scattering axis, with zero angular offset corresponding to the LOS.}
    \label{fig:brightmodel}
    \noindent\begin{minipage}{\columnwidth}
    \textbf{Alt text:} Reconstructed 1D brightness distributions of the scattered image for nine 2.25-GHz observations. The profiles vary substantially between observations, showing multiple localized components at different angular offsets around the LOS.
    \end{minipage}
\end{figure}

\begin{figure*}
    \centering
    \includegraphics[]{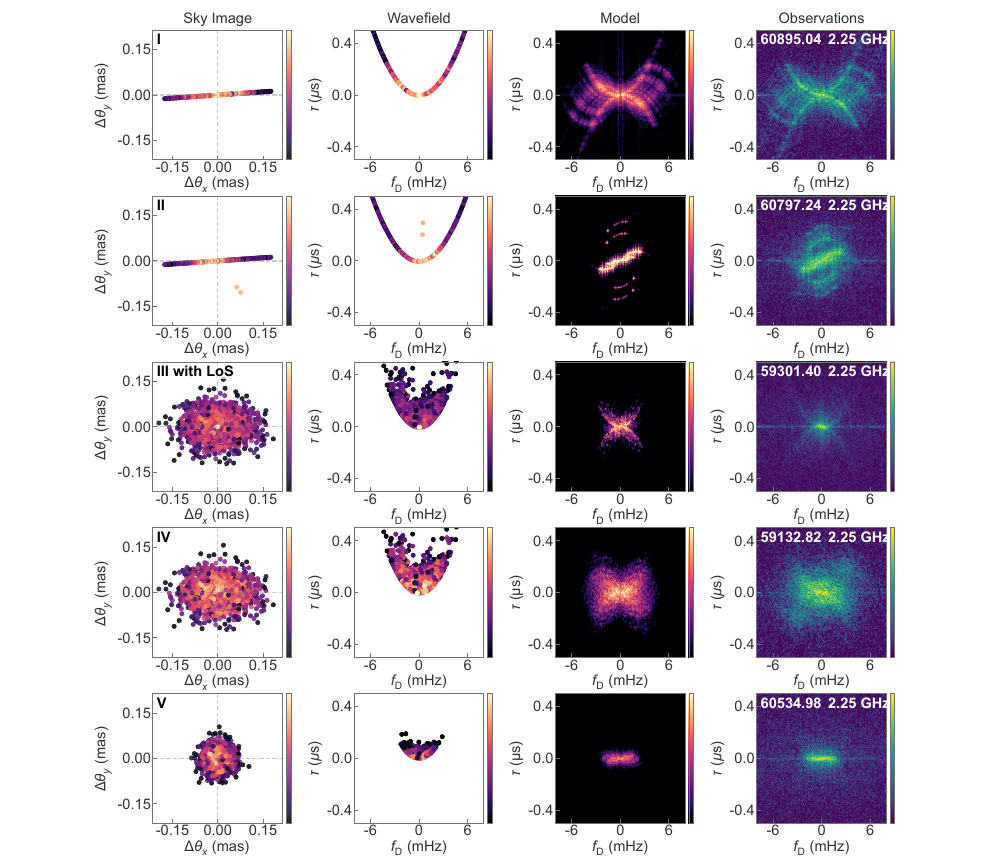}
    \caption{Simulated and observed scintillation arc morphologies. Each row shows the scattered image, wavefield, simulated secondary spectrum, and representative observation. The rows correspond to (I) a 1D image, (II) a 1D image with displaced components, (III) an extended 2D image with a pronounced component at the LOS, (IV) the same 2D spatial distribution but without a pronounced LOS component, and (V) a compact 2D image. Cases~(I) and (II) are based on reconstructions from the observations, whereas cases~(III)--(V) are qualitative simulations.}
    \label{fig:model}
    \noindent\begin{minipage}{\textwidth}
    \textbf{Alt text:} Five-row comparison of scattered images, wavefields, simulated secondary spectra, and representative observations. The 1D and displaced-component models reproduce arc and arclet structures, while extended and compact 2D images reproduce diffuse and centrally concentrated structures.
    \end{minipage}
\end{figure*}

\begin{figure*}
    \centering
    \includegraphics[width=0.9\textwidth]{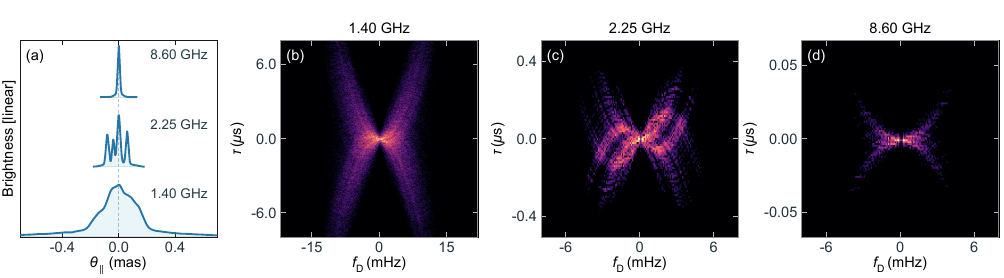}
    \caption{Simulations of representative scintillation-arc morphologies at 1.4, 2.25, and 8.60~GHz. (a) Normalized brightness profiles along the dominant scattering axis, vertically offset for clarity. (b)--(d) Corresponding simulated secondary spectra.}
    \label{fig:modelfreq}
    \noindent\begin{minipage}{\textwidth}
    \textbf{Alt text:} Four-panel comparison of representative scintillation-arc simulations at 1.4, 2.25, and 8.60~GHz. Panel (a) shows the normalized brightness profiles along the dominant scattering axis, with a broad, smooth distribution at 1.4~GHz, several localized peaks at 2.25~GHz, and a narrow, centrally concentrated distribution at 8.60~GHz. Panels (b)--(d) show the corresponding simulated secondary spectra, exhibiting a broad diffuse arc, rich substructure, and a compact forward arc, respectively.
    \end{minipage}
\end{figure*}

Within the thin-screen model, secondary spectra arise from interference between different components of the scattered image, so their morphology depends on its geometry and brightness distribution \citep{Walker2004,Cordes2006}. To explore how variations in the scattered image can reproduce the observed morphological diversity, we simulate secondary spectra using \texttt{screens}\footnote{\url{https://screens.readthedocs.io/en/latest/}} \citep{screenpy}. Each simulation adopts the same screen geometry from the anisotropic joint fit, with $D_{\mathrm{scr}}=0.931~\mathrm{kpc}$ and $\psi_0=86.06^\circ$, and the observing parameters of the corresponding observation. For highly anisotropic images that can be approximated as one-dimensional (1D), we reconstruct the angular brightness distributions along the scattering axis using the $\theta$--$\theta$ method implemented in \texttt{scintools} \citep{Reardon2020,Sprenger2021,Baker2022} and use them to guide the simulations, with representative reconstructions shown in Fig.~\ref{fig:brightmodel}. The more complex 2D distributions are illustrative constructions chosen to reproduce the main observed features.

Morphologies showing clear inverted arclets along the main arc can be reproduced with highly elongated scattered images. In these configurations, interference involving a dominant component near the LOS produces the forward arc, whereas localized components produce inverted arclets through interference with other components of the image. A representative simulation is shown in Fig.~\ref{fig:model}(I). Although the recovered brightness distributions (Fig.~\ref{fig:brightmodel}) should be regarded as approximate because of the limited reconstruction accuracy, they reveal localized structure in the scattered image on angular scales down to $\sim0.02~\mathrm{mas}$, corresponding to $0.02~\mathrm{au}$ at a screen distance of $0.931~\mathrm{kpc}$. The positions and relative brightness of the dominant components vary markedly between observations, suggesting substantial sub-au structure in the dominant scattering screen.

Displaced arclets can be reproduced by introducing compact components offset from the dominant image distribution (Fig.~\ref{fig:model}(II)). Their locations are subject to a mirror degeneracy about the direction of $\boldsymbol{V}_{\mathrm{eff}}$, so the configuration shown represents only one possible geometry. Spatially separated components within a single screen can therefore provide a viable explanation for such features \citep{Kramer2024}. For persistent features, tracking their evolution over time can provide additional constraints on their relative positions and kinematics, as discussed in Section~\ref{sec:arclet_motion}.

The 2D example in Fig.~\ref{fig:model}(III) shows that a prominent forward arc can also arise from an extended 2D image when a pronounced brightness enhancement is present at the LOS, so the forward arc alone does not uniquely determine the geometry of the scattered image. This ambiguity is also evident in previous comparisons of forward arcs produced by isotropic and anisotropic scattering \citep{Stinebring2022}. When the same extended 2D spatial configuration is considered without a pronounced LOS component, as in Fig.~\ref{fig:model}(IV), the wavefield retains a similar spatial structure, while the secondary spectrum exhibits a more diffuse arc morphology. The comparison between Fig.~\ref{fig:model}(IV) and (V) further shows that reducing the angular extent of the 2D scattered image changes the morphology from broad diffuse arcs to more compact diffuse power. Taken together, these simulations illustrate plausible scattered brightness distributions underlying the principal arc morphologies. They can also help us understand how the scattered image varies as the LOS samples different parts of the scattering structure. More irregular morphologies may reflect more complex 2D brightness distributions with localized brightness enhancements. Such cases are not modelled separately here.

The scintillation arcs observed with FAST at 1.4~GHz are generally smoother and more diffuse \citep{Ocker2024}, whereas our 2.25-GHz observations commonly show richer substructure and the 8.60-GHz arcs are typically weaker and more compact. To explore possible changes in the scattered image, we simulate representative arc morphologies at these frequencies within an anisotropic scattering framework. The corresponding brightness profiles along the dominant scattering axis are shown in Fig.~\ref{fig:modelfreq}. At 1.4~GHz, we simulate only the arc in Fig.~6 of \citet{Ocker2024} whose curvature is comparable to our measurements, using a broad, relatively smooth brightness profile without prominent localized components to represent its diffuse morphology. The broader angular extent is consistent with the larger characteristic scattering angle expected at lower frequencies \citep{Stinebring2019}. At 2.25~GHz, a narrower profile with several prominent localized peaks produces distinct arclets and richer substructure through interference between different components. At 8.60~GHz, a more compact profile dominated by a prominent component near the LOS, with much weaker outer components, produces a simple forward arc. Together, these cases suggest that the prominence of localized components may vary with frequency, making rich substructure most apparent over a limited frequency range. Since the 1.4-GHz observations were not simultaneous with our data, future  simultaneous observations at these frequencies will be important for constraining the frequency dependence of the scattered image.

\subsection{Motion and localization of compact features}
\label{sec:arclet_motion}

\begin{figure}
    \centering
    \includegraphics[]{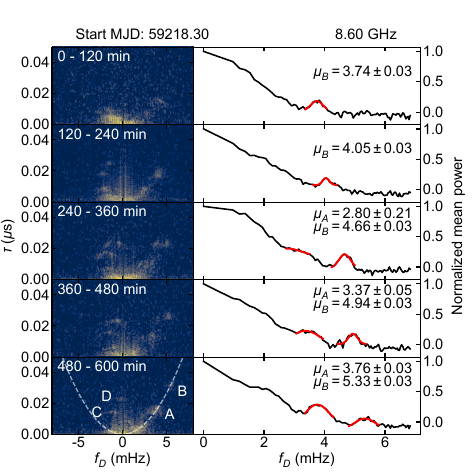}
    \caption{The five 2-h segments of the MJD~59218 observation at 8.60~GHz. Left: secondary spectra with arclets A--D marked. Right: power distributions along the main arc, with Gaussian fits to arclets A and B shown in red.}
    \label{fig:arclet_power}
    \noindent\begin{minipage}{\columnwidth}
    \textbf{Alt text:} Five consecutive 2-h segments of the 8.60-GHz observation. The left panels show secondary spectra with arclets A--D marked. The right panels show power distributions along the main arc, with red Gaussian fits to arclets A and B. The arclet positions evolve across the five segments.
    \end{minipage}
\end{figure}

\begin{table*}
    \centering
    \caption{Arclet apex measurements and derived kinematic parameters for arclets A--D.}
    \label{tab:arclet_parameters}
    \begin{threeparttable}
    \setlength{\tabcolsep}{2.5pt}
    \renewcommand{\arraystretch}{1.05}
    \begin{tabular*}{\textwidth}{@{\extracolsep{\fill}}cccccccccc@{}}
        \hline
        Arclet
        & $t$
        & $f_{\mathrm{D},g}$
        & $\tau_g$
        & $\dot{f}_{\mathrm{D},g}$
        & $V_{\mathrm{LOS}}$
        & $\eta_{\mathrm{tr}}^{\mathrm{drift}}$
        & $\tau_{\mathrm{min},g}$
        & $\eta_{\mathrm{tr}}^{\mathrm{fit}}$
        & $\left|\theta_{\perp\mathrm{V},g}\right|$ \\

        & (min)
        & (mHz)
        & ($10^{-2}\,\mu\mathrm{s}$)
        & ($10^{-3}\,\mathrm{mHz\,min^{-1}}$)
        & ($\mathrm{km\,s^{-1}}$)
        & ($10^{-3}\,\mathrm{s^3}$)
        & ($10^{-2}\,\mu\mathrm{s}$)
        & ($10^{-3}\,\mathrm{s^3}$)
        & (mas) \\
        \hline

        \multirow{5}{*}{A}
        & 60
        & --
        & --
        & \multirow{5}{*}{$3.41\pm0.44$}
        & \multirow{5}{*}{$62.7\pm8.7$}
        & \multirow{5}{*}{$1.02\pm0.13$}
        & \multirow{5}{*}{--}
        & \multirow{5}{*}{--}
        & \multirow{5}{*}{--} \\
        & 180 & -- & -- & & & & & & \\
        & 300 & $2.80\pm0.21$ & -- & & & & & & \\
        & 420 & $3.37\pm0.05$ & -- & & & & & & \\
        & 540 & $3.76\pm0.03$ & -- & & & & & & \\[1pt]

        \multirow{5}{*}{B}
        & 60
        & $3.74\pm0.03$
        & --
        & \multirow{5}{*}{$3.39\pm0.08$}
        & \multirow{5}{*}{$62.5\pm7.7$}
        & \multirow{5}{*}{$1.03\pm0.02$}
        & \multirow{5}{*}{--}
        & \multirow{5}{*}{--}
        & \multirow{5}{*}{--} \\
        & 180 & $4.05\pm0.03$ & -- & & & & & & \\
        & 300 & $4.66\pm0.03$ & -- & & & & & & \\
        & 420 & $4.94\pm0.03$ & -- & & & & & & \\
        & 540 & $5.33\pm0.03$ & -- & & & & & & \\[1pt]

        \multirow{5}{*}{C}
        & 60
        & --
        & --
        & \multirow{5}{*}{$3.04\pm0.46$}
        & \multirow{5}{*}{$59.2\pm8.5$}
        & \multirow{5}{*}{$1.15\pm0.17$}
        & \multirow{5}{*}{$1.34\pm0.05$}
        & \multirow{5}{*}{$1.00\pm0.18$}
        & \multirow{5}{*}{$0.029\pm0.004$} \\
        & 180 & $-2.21\pm0.11$ & $1.83\pm0.02$ & & & & & & \\
        & 300 & $-1.84\pm0.14$ & $1.67\pm0.03$ & & & & & & \\
        & 420 & $-1.50\pm0.11$ & $1.57\pm0.01$ & & & & & & \\
        & 540 & $-1.10\pm0.14$ & $1.46\pm0.03$ & & & & & & \\[1pt]

        \multirow{5}{*}{D}
        & 60
        & --
        & --
        & \multirow{5}{*}{$3.15\pm0.63$}
        & \multirow{5}{*}{$60.2\pm9.5$}
        & \multirow{5}{*}{$1.11\pm0.22$}
        & \multirow{5}{*}{$2.28\pm0.02$}
        & \multirow{5}{*}{$0.93\pm0.28$}
        & \multirow{5}{*}{$0.037\pm0.005$} \\
        & 180 & $-1.55\pm0.15$ & $2.51\pm0.04$ & & & & & & \\
        & 300 & $-0.97\pm0.15$ & $2.36\pm0.04$ & & & & & & \\
        & 420 & $-0.72\pm0.11$ & $2.33\pm0.02$ & & & & & & \\
        & 540 & $-0.42\pm0.23$ & $2.30\pm0.02$ & & & & & & \\

        \hline
    \end{tabular*}
    \end{threeparttable}
    \vspace{4pt}
    \parbox{\textwidth}{\small\textit{Note.} $t$ denotes the elapsed time at the centre of each segment. For arclets A and B, $\tau_g$ is not independently measured. The uncertainties in $V_{\mathrm{LOS}}$ and $\left|\theta_{\perp\mathrm{V},g}\right|$ include those in $s$ and $D_p$.}
\end{table*}

\begin{figure*}
    \centering
    \includegraphics[]{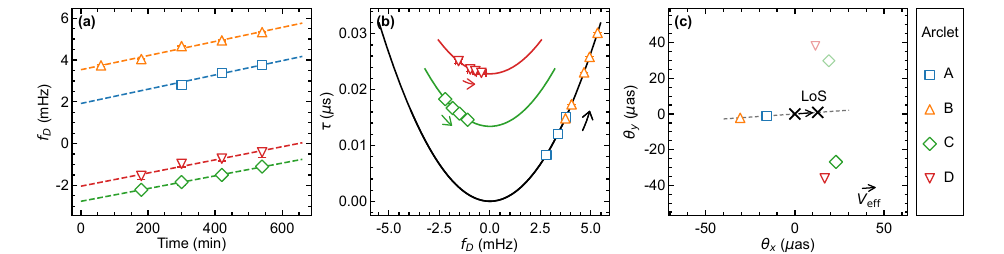}
    \caption{Evolution of the four arclets (A--D).
    (a) Temporal evolution of $f_{\mathrm{D},g}$ and the corresponding linear fits.
    (b) Trajectories of the arclet apices in the secondary spectrum. The arrows indicate the direction of evolution.
    (c) Reconstructed sky-plane positions of the four scattered images for the limiting case $\Delta=0$, referenced to the LOS position in the first segment. The dashed line shows the main scattering axis and the arrow traces the LOS displacement to the fifth segment, with faint symbols denoting the mirrored solutions.}
    \label{fig:arclet_move}
    \noindent\begin{minipage}{\textwidth}
    \textbf{Alt text:} Three-panel representation of the evolution of arclets A--D. Their Doppler coordinates vary approximately linearly with time, while the measured apex trajectories are consistent with parabolas symmetric about $f_{\mathrm D}=0$. For the limiting case $\Delta=0$, the reconstructed sky-plane positions show that the scattered images associated with arclets A and B lie along the dominant scattering axis, whereas those associated with C and D are displaced from this axis.
    \end{minipage}
\end{figure*}

In the 10-h observation at 8.60~GHz on MJD 59218, four arclets, labelled A--D in Fig.~\ref{fig:arclet_power}, were identified and tracked over multiple consecutive 2-h segments. Tracking the temporal evolution of their apices provides constraints on the relative geometry between the LOS and the corresponding localized scattered images. Related work on arclet evolution in 1.4-GHz observations of PSR~B0355+54 has been carried out by Yao et al. (in preparation).

Over the 10-h observation, variations in $D_{\mathrm{eff}}$ and $\boldsymbol{V}_{\mathrm{eff}}$ are negligible. Inverted arclets A and B lie on the main arc and are associated with localized scattered images along the dominant scattering axis. For the displaced arclets C and D, the corresponding localized scattered images can be treated as approximately stationary relative to the scattering screen while the LOS moves across them. The corresponding geometric relations are derived in Appendix~\ref{app:geometric_arc}. Following previous studies \citep{Hill2005,Wang2018}, the arclet apices were measured from their power distributions in the secondary spectra. For inverted arclets A and B, the Doppler coordinates were determined by Gaussian fits to the power peaks along the main arc. For the displaced arclets C and D, both apex coordinates $(f_{\mathrm{D},g},\tau_g)$ were determined by sliding an inverted parabolic template across the secondary spectrum and locating the position of maximum integrated power. The results are listed in Table~\ref{tab:arclet_parameters}.

As shown in Fig.~\ref{fig:arclet_move}(a) and (b), the Doppler coordinates of all four arclets vary approximately linearly with time, with consistent drift rates of $3.0$--$3.4\times10^{-3}~\mathrm{mHz~min^{-1}}$. For the displaced arclets C and D, the measured apex evolution is well described by equation~\eqref{eq:arclet_apex_trajectory}, and the fitted trajectory curvatures $\eta_{\mathrm{tr}}^{\mathrm{fit}}$ are consistent with the corresponding values $\eta_{\mathrm{tr}}^{\mathrm{drift}}$ inferred from their Doppler drift rates. If the displaced arclets C and D arise from the same dominant scattering screen as the main arc, then $\eta_{\mathrm{tr}}=\eta\cos^{2}\Delta$. Their trajectory curvatures agree within uncertainties with the contemporaneous main arc curvature ($(1.06\pm0.05)\times10^{-3}~\mathrm{s}^{3}$), suggesting that the angle between $\boldsymbol{V}_{\mathrm{eff}}$ and the dominant scattering axis may be small.

In this case, the similar Doppler drift rates of the four arclets correspond to a transverse LOS sweep speed of $V_{\mathrm{LOS}}=sV_{\mathrm{eff}}\simeq61~\mathrm{km~s^{-1}}$, or an angular velocity of about $14~\mathrm{mas~yr^{-1}}$. Such motion implies that the LOS can shift by an angular distance comparable to the $0.3~\mathrm{mas}$ extent of the scattered image in about eight days, which may account for the morphological changes observed on time-scales of days.

The non-zero minimum apex delays of the displaced arclets C and D give $|\theta_{\perp\mathrm{V},g}|=0.029\pm0.004$ and $0.037\pm0.005~\mathrm{mas}$, corresponding to projected offsets of $0.027$ and $0.034~\mathrm{au}$ from the LOS in the direction perpendicular to $\boldsymbol{V}_{\mathrm{eff}}$. For the limiting case $\Delta=0$, we reconstruct the sky-plane positions of the four scattered images and the LOS motion in Fig.~\ref{fig:arclet_move}(c). The reconstruction is not unique because the sign of $\theta_{\perp\mathrm{V},g}$ is unconstrained, giving mirrored solutions. However, the intrinsic angular sizes of these scattered image components remain poorly constrained by the current 8.60-GHz data.

\subsection{Implications of arclet kinematics for the screen velocity}
\label{sec:screenv}
The annual curvature fit provides only weak constraints on the full transverse screen velocity in the anisotropic model. We therefore consider the limiting case $\Delta=0$ to examine its implications for the screen velocity.

At MJD~59218, the transverse velocities of the pulsar and Earth are $\boldsymbol{V}_{\mathrm{p}}=(43.61,38.72)$ and $\boldsymbol{V}_{\oplus}=(22.15,14.96)~\mathrm{km~s^{-1}}$, respectively, in the directions of increasing right ascension and declination. With $\hat{\boldsymbol{e}}_{\parallel}=(\sin\psi_0,\cos\psi_0)$, the screen velocity is
\begin{equation}
    \boldsymbol{V}_{\mathrm{scr}}
    =
    (1-s)\boldsymbol{V}_{\mathrm{p}}
    +
    s\boldsymbol{V}_{\oplus}
    -
    \epsilon V_{\mathrm{LOS}}
    \hat{\boldsymbol{e}}_{\parallel},
    \label{eq:screen_velocity}
\end{equation}
where $\epsilon=\pm1$ accounts for the unresolved direction along the scattering axis.

Using $s=0.069$, $\psi_0=86.06^\circ$, and $V_{\mathrm{LOS}}\simeq61~\mathrm{km~s^{-1}}$ gives two possible screen velocities, $\boldsymbol{V}_{\mathrm{scr}}\simeq(-19,33)$ and $(103,41)~\mathrm{km~s^{-1}}$. Of the two solutions, $\boldsymbol{V}_{\mathrm{scr}}\simeq(-19,33)~\mathrm{km~s^{-1}}$ is closer to the screen velocity estimate from the annual curvature fit. However, this estimate remains conditional on the assumption $\Delta=0$ and on the screen parameters obtained from the annual analysis. More precise constraints on $\Delta$ and arclet evolution measurements from multiple observations could be incorporated directly into the annual curvature analysis, allowing the screen velocity and geometry to be determined more accurately.

\subsection{Comparison with previous scintillation studies of PSR~B0355+54}
\label{sec:Comparison}
Compared with previous 2.25-GHz studies, our TMRT observations reveal a broader range of clearly resolved arc morphologies, which may partly reflect differences in observational sensitivity and spatial variations within the scattering region sampled by the LOS over time. Despite this diversity, the measured scaled arc curvatures fall within $223$--$346~\mathrm{m^{-1}\,mHz^{-2}}$, comparable to the value measured with JM66 ($\sim365~\mathrm{m^{-1}\,mHz^{-2}}$) \citep{Wang2018} and to one of the prominent arcs detected with FAST ($\sim327~\mathrm{m^{-1}\,mHz^{-2}}$) \citep{Ocker2024}. The annual modulation of the arc curvature places the dominant scattering screen at $s=0.069^{+0.015}_{-0.014}$, close to the pulsar and consistent with both the JM66 estimate of $s\simeq0.081$ and the near-pulsar solution ($0.031\lesssim s\lesssim0.08$) inferred from the FAST measurement with a similar curvature. This localization provides further evidence for a possible association of the dominant screen with the \ion{H}{ii} region Sh~2-205.

FAST observations at 1.4~GHz revealed several arcs with distinct curvatures toward PSR~B0355+54 \citep{Ocker2024}. Most of our TMRT observations instead show a single clear arc, although a few contain additional structures for which it is unclear whether they belong to the same arc or represent separate arcs, such as the central structure in panel (b) of Fig.~\ref{fig:sspec}. This difference from the FAST results may reflect variations in the relative prominence of different arcs with observing frequency, with differences in observational sensitivity also affecting their detectability. Alternatively, the relative contributions of different scattering screens may change as the LOS moves across the scattering medium. The large curvature variations measured at 2.25~GHz with KM40 ($39$--$241~\mathrm{m^{-1}\,mHz^{-2}}$) are consistent with the latter possibility \citep{Xu2018}. The continued detection of the same arc throughout our long-term monitoring further suggests that the associated screen is a long-lived and dominant component of the scattering environment along this LOS.

Previous work tracked inverted arclets along the main arc of PSR~B0355+54 at 2.25~GHz \citep{Wang2018}. Our 8.60-GHz observation extends this analysis to displaced arclets, providing additional constraints on the scattering geometry from their apex evolution. The measurements suggest that the direction of $\boldsymbol{V}_{\mathrm{eff}}$ may be close to the dominant scattering axis. The similar Doppler drift rates of all four arclets then yield an estimated LOS angular velocity of $\sim14~\mathrm{mas\,yr^{-1}}$, consistent with the previously reported values of $14\pm5$ and $18\pm7~\mathrm{mas\,yr^{-1}}$ \citep{Wang2018}.

Although most of our results are consistent with a dominant scattering screen, contributions from additional screens or scattering paths cannot be excluded \citep{Xu2018,Ocker2024,Zhu2023,Chen2025}. Simultaneous ultra-wideband observations could better trace the frequency evolution of individual scintillation arcs and help distinguish contributions from multiple screens. In addition, intensity secondary spectra cannot uniquely determine the full 2D scattered image, and uncertainties in arclet apex positions further limit constraints on $\Delta$. Higher-resolution scintillation observations with spatial baseline information, together with studies of the intervening ISM, would help constrain the scattering geometry and kinematics more precisely and establish the physical origin of the scattering structures.

\section{Conclusions}
\label{sec:5}
Our long-term simultaneous 2.25/8.60~GHz observations reveal for the first time that PSR~B0355+54 exhibits nearly the full range of known scintillation arc morphologies, many of which evolve rapidly on time-scales from days to hours. They also uncover a clear frequency dependence, with rich and highly variable arc structures at 2.25~GHz compared with much simpler and fainter arcs at 8.60~GHz. Despite these pronounced morphological differences, the scaled curvatures and their annual variations are consistent between the two frequencies, indicating a common dominant scattering screen. The joint annual curvature fit robustly localizes this screen at a fractional distance of $s=0.069^{+0.015}_{-0.014}$ near the pulsar, further supporting a possible association with the \ion{H}{ii} region Sh~2-205. The ACF-derived pseudo-curvatures show broad agreement with the measured arc curvatures, providing an additional consistency check on this interpretation. Our simulations show that a wide range of the observed arc morphologies can arise from complex variations in the scattered-image brightness distribution within this dominant screen geometry, with the 2.25-GHz reconstructions revealing localized components on angular scales down to $\sim0.02~\mathrm{mas}$ ($0.02~\mathrm{au}$). The frequency-dependent models further suggest that rich substructure may be most apparent over a limited frequency range, depending on the relative prominence of localized brightness components. These results substantially broaden the observational picture established by previous studies of PSR~B0355+54 and indicate pronounced small-scale variations in the scattering properties of the ionized medium along this LOS.

The 10-h observation at 8.60~GHz extends previous arclet-tracking studies of PSR~B0355+54 to displaced arclets, whose apex evolution provides additional constraints on the angular positions of the associated scattered components and on the direction of $\boldsymbol{V}_{\mathrm{eff}}$ relative to the dominant scattering axis. The measurements suggest that the direction of $\boldsymbol{V}_{\mathrm{eff}}$ may be close to the dominant scattering axis and imply that the LOS sweeps across the scattering screen at an angular rate of $\sim14~\mathrm{mas~yr^{-1}}$. At this rate, the LOS shifts by about $0.3~\mathrm{mas}$ in eight days, comparable to the extent of the scattered image reconstructed at 2.25~GHz, providing a natural explanation for the observed day-scale changes in arc morphology. These results make future higher-precision observations with broader simultaneous frequency coverage particularly exciting for tracing the frequency evolution of scintillation arcs and further constraining the geometry, kinematics, and small-scale structure of the scattering screen.

\section*{Acknowledgements}
This work is supported in part by the National Key R$\&$D Program of China (grant No. 2022YFA1603104), the National SKA Program of China (grant No. 2020SKA0120104), the National Natural Science Foundation of China (grant Nos. 12041301, U2031119), and the Natural Science Foundation of Shanghai (grant No. 20ZR1467600). Generative AI tools were used to assist with language editing and code used in figure preparation. All scientific content, analyses, numerical results, figures, interpretations, and the final manuscript were checked and approved by the authors, who take full responsibility for the work.

\section*{Data Availability}
The data underlying this article will be shared on reasonable request to the corresponding authors.

\bibliographystyle{mnras}
\bibliography{reference}


\appendix

\section{Arc morphology classification}
\label{app:morphology}
To distinguish the different types of arc structures and morphological categories used throughout this paper, the relevant terminology is summarized below.

\textbf{Main arcs:} Standard forward parabolic structures with their apices located at the origin of the secondary spectrum.

\textbf{Offset forward arcs:} Forward parabolic structures whose apices are offset from the origin.

\textbf{Inverted arclets:} Inverted parabolic features whose apices lie on the main arc.

\textbf{Displaced arclets:} Inverted parabolic features whose apices are displaced from the main arc.

\textbf{Diffuse arcs:} Broad power distributions that retain an overall parabolic shape or contain a few faint arclets.

\textbf{Diffuse power:} Diffuse power with no identifiable shape or distinct arc structures.

\textbf{Irregular structures:} Highly distorted, asymmetric, or fragmented power distributions without a coherent overall arc geometry.

The secondary spectra are grouped into five broad empirical categories according to their dominant overall morphology: (M1) forward arcs, including both main and offset forward arcs; (M2) inverted arclets; (M3) diffuse arcs; (M4) diffuse power; and (M5) irregular structures.  The symbol $(\star)$ marks observations that additionally contain prominent displaced arclets.

\section{Motion of displaced arclets}
\label{app:geometric_arc}
For a persistent displaced arclet, its apex corresponds to interference between a localized scattered image at angular position $\boldsymbol{\theta}_g$ relative to the LOS and the reference image at $\boldsymbol{\theta}=0$. Over the tracking period, $\boldsymbol{V}_{\mathrm{eff}}$ is assumed to be constant. If the localized image remains stationary relative to the screen, the motion of the LOS only changes the component of $\boldsymbol{\theta}_g$ parallel to $\boldsymbol{V}_{\mathrm{eff}}$, while the perpendicular component remains constant. Decomposing $\boldsymbol{\theta}_g$ into these two components gives
\begin{equation}
    f_{\mathrm{D},g}
    =
    -\frac{\nu V_{\mathrm{eff}}}{c}\theta_{\parallel\mathrm{V},g},
    \label{eq:apex_fd_v}
\end{equation}
and
\begin{equation}
    \tau_g
    =
    \frac{D_{\mathrm{eff}}}{2c}
    \left(
    \theta_{\parallel\mathrm{V},g}^{2}
    +
    \theta_{\perp\mathrm{V},g}^{2}
    \right).
    \label{eq:apex_tau_v}
\end{equation}
The apex therefore follows
\begin{equation}
    \tau_g
    =
    \tau_{\min,g}
    +
    \eta_{\mathrm{tr}}f_{\mathrm{D},g}^{2},
    \label{eq:arclet_apex_trajectory}
\end{equation}
where $\tau_{\min,g}=D_{\mathrm{eff}}\theta_{\perp\mathrm{V},g}^{2}/(2c)$ and
$\eta_{\mathrm{tr}}=cD_{\mathrm{eff}}/(2\nu^{2}V_{\mathrm{eff}}^{2})$.

If the localized image is associated with the same thin screen as the main arc, its trajectory curvature satisfies
\begin{equation}
    \eta_{\mathrm{tr}}=\eta\cos^{2}\Delta,
    \label{eq:arclet_trajectory_curvature}
\end{equation}
where $\Delta$ is the angle between $\boldsymbol{V}_{\mathrm{eff}}$ and the dominant scattering axis. Comparison with the main arc curvature $\eta$ therefore constrains $|\Delta|$. The corresponding Doppler drift rate is given by \citet{Shi2021} as
\begin{equation}
    \left|\dot{f}_{\mathrm{D},g}\right|
    =
    \frac{\nu V_{\mathrm{eff}}^{2}}
    {cD_{\mathrm{eff}}},
    \label{eq:arclet_doppler_rate}
\end{equation}
which gives $\eta_{\mathrm{tr}}=1/(2\nu|\dot{f}_{\mathrm{D},g}|)$.

The minimum apex delay gives the magnitude of the component of $\boldsymbol{\theta}_g$ perpendicular to the direction of $\boldsymbol{V}_{\mathrm{eff}}$
\begin{equation}
    \left|\theta_{\perp\mathrm{V},g}\right|
    =
    \left(
    \frac{2c\tau_{\min,g}}{D_{\mathrm{eff}}}
    \right)^{1/2}.
    \label{eq:arclet_perp_offset}
\end{equation}


\bsp
\label{lastpage}
\end{document}